\documentclass{spie}
\usepackage{amsmath}
\usepackage{amssymb}
\usepackage{fontawesome}
\usepackage{graphicx}
\usepackage[version=4]{mhchem}
\usepackage[binary-units]{siunitx}
\usepackage[caption=false]{subfig}
\usepackage{units}
\usepackage{todonotes}
\usepackage[colorlinks=true, allcolors=blue]{hyperref}
\pdfmapfile{=fontawesome.map}

\DeclareSIUnit\parsec{pc}

\newcommand{\HeIII}{\ce{^{3}}He}
\newcommand{\HeIV}{\ce{^{4}}He}

\title{MUSCAT: The Mexico-UK Sub-Millimetre Camera for AsTronomy}

\author[1]{Thomas L. R. Brien}
\author[1]{Peter A. R. Ade}
\author[1,2]{Peter S. Barry}
\author[3]{Edgar Castillo-Dom\`{i}nguez}
\author[3]{Daniel Ferrusca}
\author[1]{Thomas Gascard}
\author[3]{Victor G\'{o}mez}
\author[1]{Peter C. Hargrave}
\author[1]{Amber L. Hornsby}
\author[3]{David Hughes}
\author[4]{Enzo Pascale}
\author[1]{Josie D. A. Parrianen}
\author[3]{Abel Perez}
\author[1]{Sam Rowe}
\author[1]{Carole Tucker}
\author[3]{Salvador Ventura Gonz\'{a}lez}
\author[1]{Simon M. Doyle}

\affil[1]{School of Physics \& Astronomy, Cardiff University, The Parade, Cardiff, CF24 3AA, United Kingdom}
\affil[2]{Kavli Institute for Cosmological Physics, University of Chicago, Chicago, Illinois, United States of America}
\affil[3]{Instituto Nacional de Astrof\'{i}sica, \'{O}ptica y Electr\'{o}nica, Luis Enrique Erro 1, Santa Mar\'{i}a Tonatzintla, 72840 Puebla, Mexico}
\affil[4]{Dipartimiento di Fisica, La Sapienza Universit\`{a} di Roma, Piazzale Aldo Moro, 5, 00185 Roma, Italy}

\authorinfo{Corresponding author: Thomas L. R. Brien (\faicon{envelope-o} \href{mailto:tom.brien@astro.cf.ac.uk}{tom.brien@astro.cf.ac.uk} \faicon{phone} \href{tel:+442920876269}{+44 (0) 29 20876269})}

\begin{document}
\maketitle

\begin{abstract}
The Mexico-UK Sub-millimetre Camera for AsTronomy (MUSCAT) is a large-format, millimetre-wave camera consisting of 1,500 background-limited lumped-element kinetic inductance detectors (LEKIDs) scheduled for deployment on the Large Millimeter Telescope (Volc\'{a}n Sierra Negra, Mexico) in 2018. MUSCAT is designed for observing at 1.1 mm and will utilise the full $4\si{\arcminute}$ field of view of the LMTs upgraded 50-m primary mirror. In its primary role, MUSCAT is designed for high-resolution follow-up surveys of both galactic and extra-galactic sub-mm sources identified by \textit{Herschel}. MUSCAT is also designed to be a technology demonstrator that will provide the first on-sky demonstrations of novel design concepts such as horn-coupled LEKID arrays and closed continuous cycle miniature dilution refrigeration.
\par
Here we describe some of the key design elements of the MUSCAT instrument such as the novel use of continuous sorption refrigerators and a miniature dilutor for continuous 100-mK cooling of the focal plane, broadband optical coupling to Aluminium LEKID arrays using waveguide chokes and anti-reflection coating materials as well as with the general mechanical and optical design of MUSCAT. We explain how MUSCAT is designed to be simple to upgrade and the possibilities for changing the focal plane unit that allows MUSCAT to act as a demonstrator for other novel technologies such as multi-chroic polarisation sensitive pixels and on-chip spectrometry in the future. Finally, we will report on the current status of MUSCAT's commissioning.
\end{abstract}

\section{Introduction}
The Mexico-UK Sub-mm Camera for AsTronomy (MUSCAT) is a receiver currently being developed for the Large Millimeter Telescope (LMT; Spanish: Gran Telescopio Milim\'{e}trico, GTM) on Volc\'{a}n Sierra Negra in Puebla Mexico. In it's first generation MUSCAT will consist of a focal plane of approximately 1,500 lumped-element kinetic inductance detectors (LEKIDs) and will have a single operating band centred about $1.1~\si{\milli\metre}$. MUSCAT utilises solely reflective optical components to couple to the tertiary mirror of the LMT.
\par 
During it's first semester on the LMT, MUSCAT will be used to explore the evolution of galaxies and to map star-forming regions beyond the Gould belt. Due to confusion limits arising from the resolution of currently available instruments, approximately two thirds of the sources given in the \textit{Herschel}-ATLAS catalogue do not have optical counterparts allocated.\cite{Furlanetto2018}. With its much greater resolution, MUSCAT will help to assign optical counterparts for these sources. MUSCAT will also be used to study the link between the evolution of filament structures and star formation. Currently the study of this relationship has been resolution limited to studying the Gould belt star-forming clouds ($d < 400~\si{\parsec}$). MUSCAT on the LMT will be able to map star-forming regions out to $d = 4~\si{\kilo\parsec}$.
\par 
The first-generation specification and design of MUSCAT has been finalised and the instrument key features of this design include continuous cooling of the focal plane to of order $100~\si{\milli\kelvin}$, the use of an anti-reflection coating across the focal plane, and the use of waveguide-chokes to define broadband optical coupling to the detector array. The mechanical and optical design of MUSCAT is designed such that it is possible to upgrade MUSCAT with minimal complexity. As such, beyond its first semester, it is envisioned that MUSCAT will be able to be used as an on-sky demonstrator for emerging technologies in sub-mm instrumentation. Such technologies currently considered include on-chip spectrometry,  band-defined detectors using on-chip band-pass filters, and also higher MUX ratio readouts.
\section{The Large Millimeter Telescope}
The Large Millimeter Telescope Alfonso Serrano is currently the world's largest single-dish millimetre-wave telescope having recently completed an upgrade of the primary surface to a $50~\si{\metre}$ diameter. This primary surface consists of five concentric rings of reflective panels with a total of 180 panels in total. Each of these individual panels is actively controlled via actuators with further active control of the secondary surface via a hexapod. The surface roughness of the primary is between $16\mbox{--}16~\si{\micro\metre}$ r.m.s., with the panels in the more recently installed outer two rings having the lower surface roughness. The secondary mirror has a r.m.s. surface finish of $30~\si{\micro\metre}$
\par
\begin{figure}[tb]
\centering
\includegraphics[height=0.45\textheight]{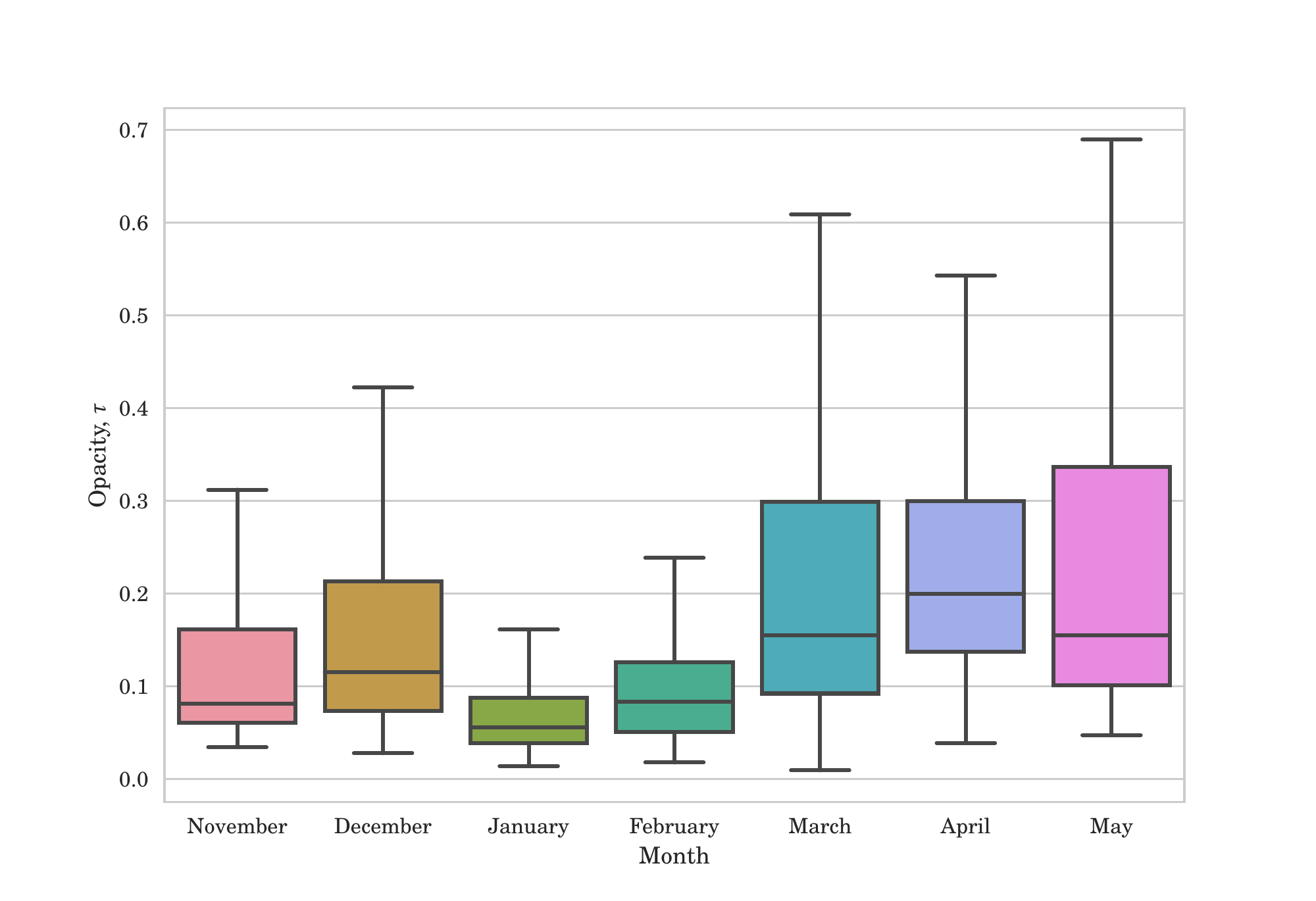}
\caption{Tuckey box and whisker plot of the opacity at the LMT during nighttime (10 PM to 6 AM) in the 2016/17 winter observing season.}\label{fig:tau}
\end{figure}

 
The LMT is located at an altitude of $4,600~\si{\metre}$ at the summit of Volc\'{a}n Sierra Negra in Puebla State, Mexico. During the November 2016--May 2017 observing season, the LMT enjoyed median nighttime observing conditions of $\tau < 0.2$, as seen in Fig.~\ref{fig:tau}. The conditions at the LMT site are described fully by Ferrusca and Contreras,\cite{Ferrusca2014} and Zeballos et al.\cite{Zeballos2016}
\section{Instrument Vessel Design}
To allow for flexibility and the possibility of future upgrades, the MUSCAT vacuum vessel is designed to be similar to many test cryostats and consists of a nested set of radiation shields within the vacuum shield, all of which assemble like a set of Russian Matryoshka dolls. The outer vacuum shield is $900~\si{\milli\metre}$ in height with an internal diameter of $840~\si{\milli\metre}$. The inside of the vacuum is lined with FINEMET to provide magnetic shielding and a top (inner most) layer of low-emissivity material to reduce the radiation load on the subsequent stages. Within the outer vacuum vessel, three copper radiation shields are nested within each other and are mounted from the 50-, 4-, and 0.45-kelvin stages of MUSCAT (see Fig.~\ref{fig:coolingChain}). The final $450\mbox{-}\si{\milli\kelvin}$ enclosure has a diameter of $600~\si{\milli\metre}$ and is $450~\si{\milli\metre}$ in height. The $100\mbox{-}\si{\milli\kelvin}$ focal plane stage does not incorporate a full shield but does utilise a baffle shield around the focal plane to protect from stray light. This open design, along with the simple construction of the instrument vessel, should enable simple switching and upgrading of the focal plane. Further stray light protection is offered by an additional set of baffles incorporated into the optical aperture of each fully shielded stage of MUSCAT and also by the inner surface of the $450\mbox{-}\si{\milli\kelvin}$ shield which is blackened and includes a scattering material.
\begin{figure}[tb]
\centering
\includegraphics[height=0.45\textheight]{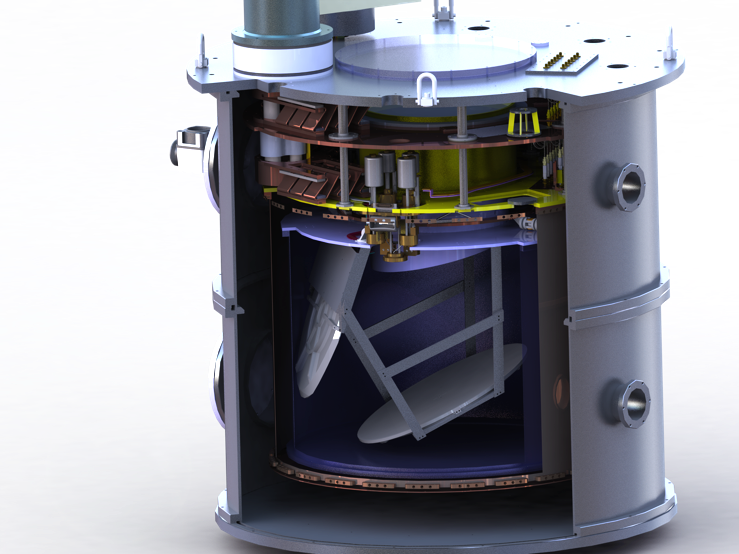}
\caption{Rendered CAD model of the final MUSCAT design. The $450\mbox{-}\si{\milli\kelvin}$ continuous sorption refrigerator can be seen towards the centre of the figure and the final two MUSCAT mirrors (M7 and M8) are shown inside the $450\mbox{-}\si{\milli\kelvin}$ shield (coloured blue).}\label{fig:CAD}
\end{figure}
\par 
The detector array is protected from mechanical vibrations from the pulse tube cooler by a series of counter measures. The pulse tube cooler is mounted to the top plate of the outer vacuum vessel via a set of vibration dampening rubber gaskets which decouple the two flanges while still maintaining a hermetic seal. The 50- and 4-kelvin heads of the pulse tube cooler are isolated from the mechanical structure of MUSCAT via the use of oxygen-free, high-purity copper braid; this  mechanically decouples these stages of MUSCAT from the pulse tube cooler while still providing sufficient thermal conduction to allow the stages to operate at an acceptable temperature under thermal load.
\par 
The $1\mbox{-}\si{\kelvin}$ stage is supported through by thin-walled stainless steel crossbeams; lab testing of these has shown that the total thermal load on this stage from these supports is approximately $7~\si{\micro\watt}$ per support.\cite{Brien2018} The $450\mbox{-}$ and $100\mbox{-}\si{\milli\kelvin}$ stages of MUSCAT are supported through by joints containing a crushed-sapphire interface which thermally load the $450\mbox{-}$ and $100\mbox{-}\si{\milli\kelvin}$ stages by $2~\si{\micro\watt}$ and $0.3~\si{\micro\watt}$ per joint respectively.\cite{Brien2018}

\section{Optical Design}
\begin{figure}[b]
\centering
\subfloat[]{
\includegraphics[height=0.2\textheight]{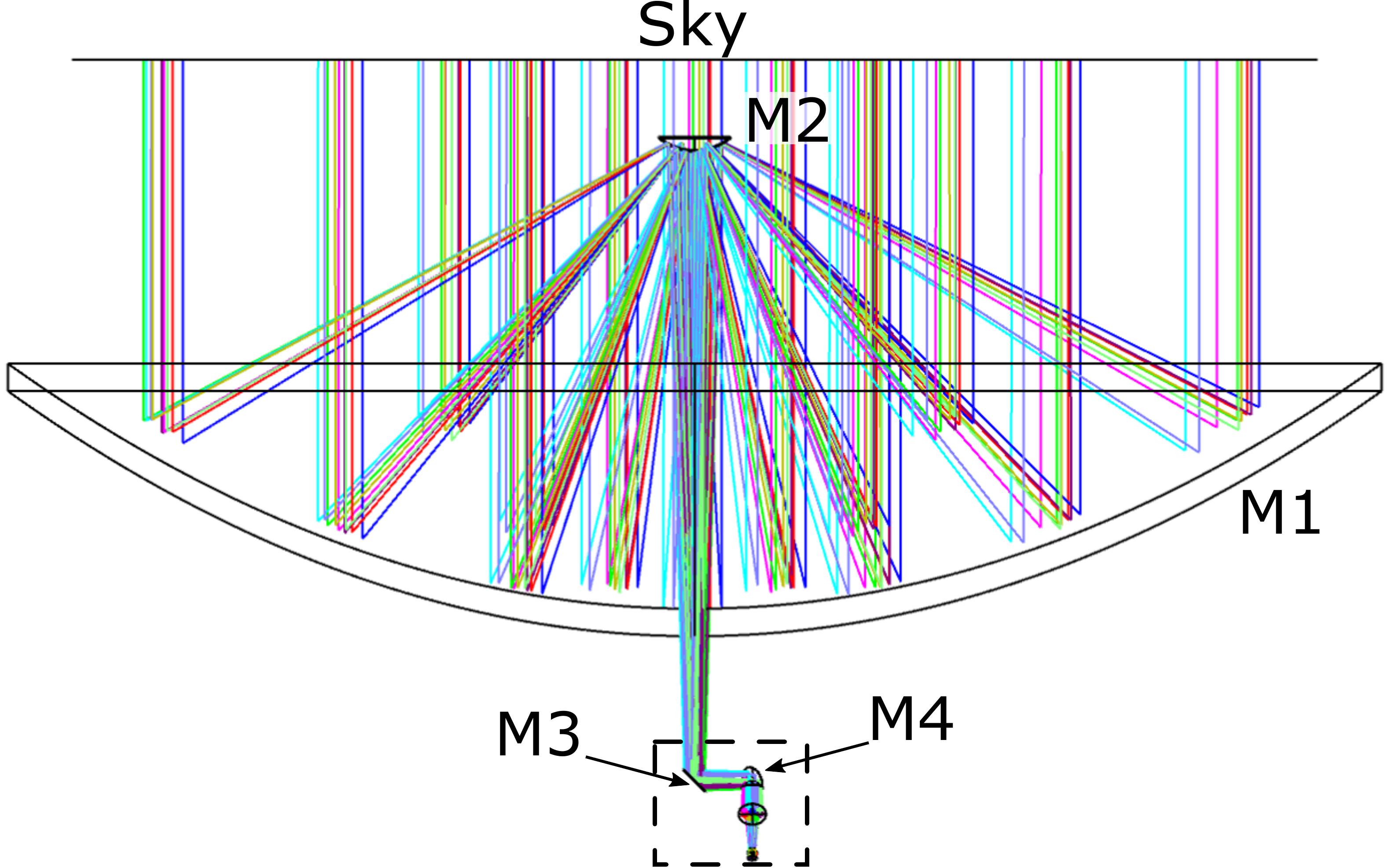}
\label{fig:RaysFull}}
\subfloat[]{
\includegraphics[height=0.2\textheight]{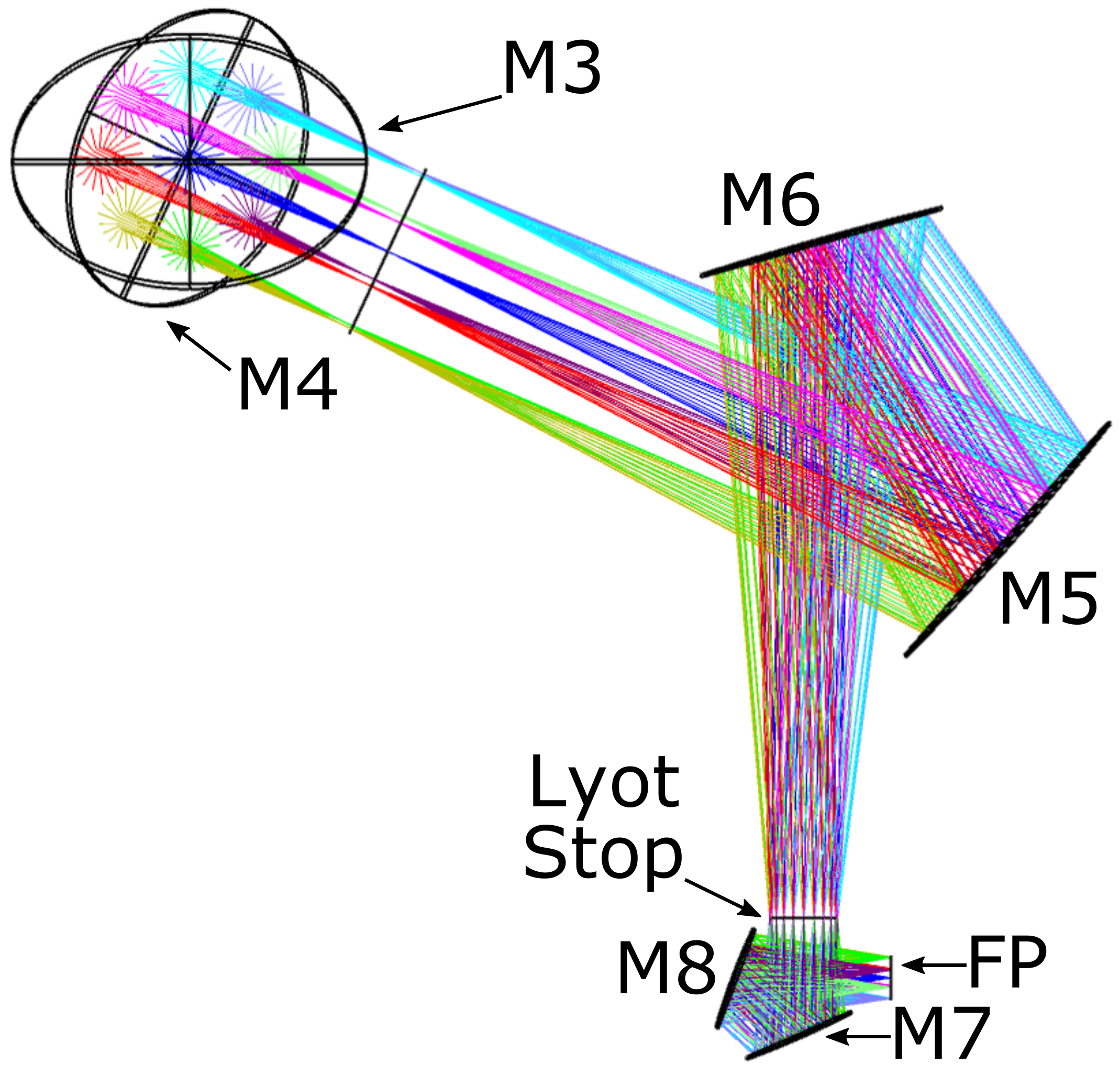}
\label{fig:RaysM3}}
\hspace{0.1\textwidth}
\subfloat[]{
\includegraphics[height=0.2\textheight]{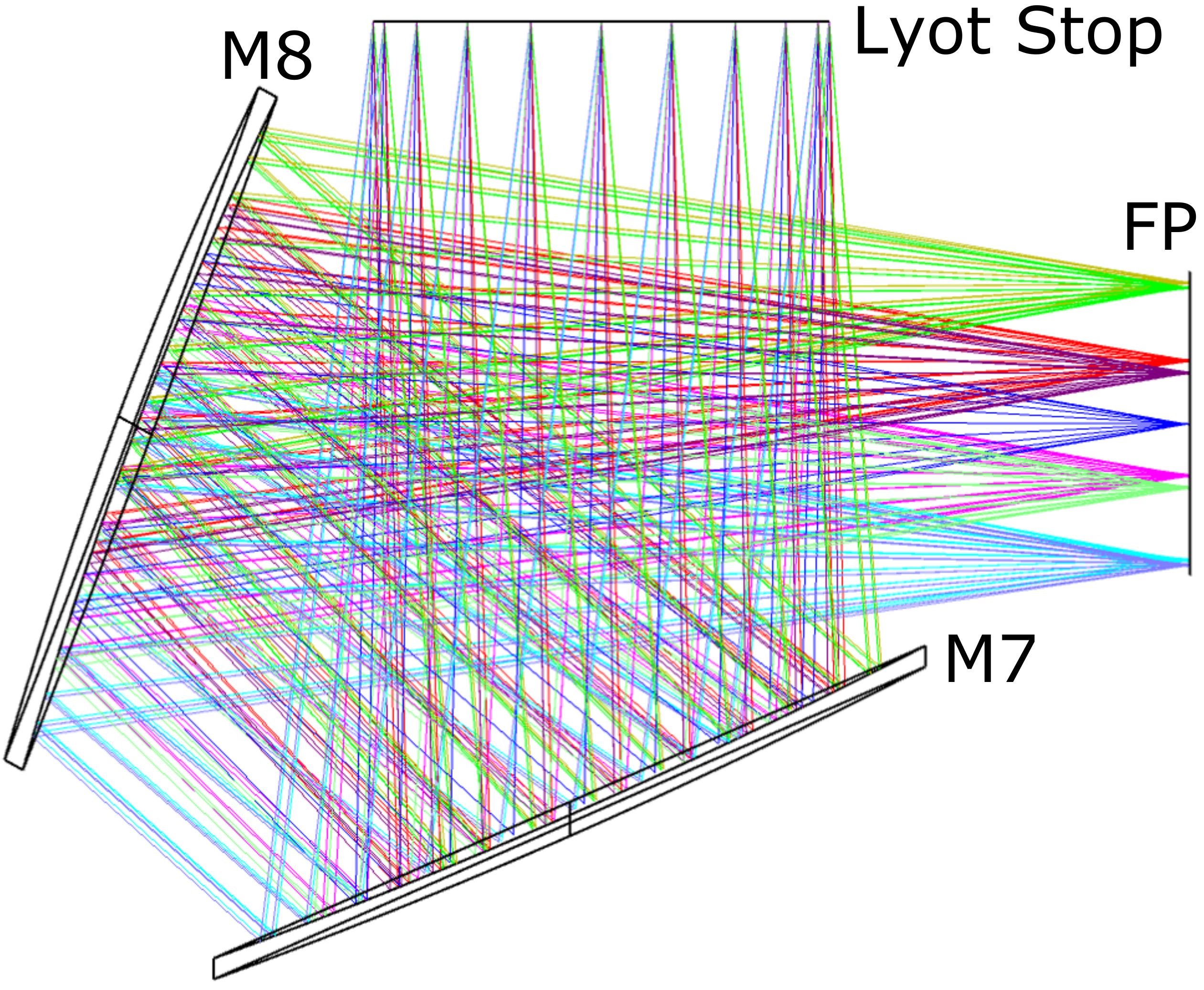}
\label{fig:RaysLyot}}
\caption{(a) Full ray trace of MUSCAT on the 50-m LMT, the dash box contains the tertiary mirror of the LMT and MUSCAT's own optics. (b) Zoomed and rotated view of the mirrors contained in the dashed area of (a), the image is rotated such that the rays between M3 and M4 are drawn perpendicular to the plane of the page. (c) Zoomed view of the cooled mirrors inside MUSCAT's cryostat with the focal plane (FP) shown to the right.}\label{fig:Rays}
\end{figure}

The LMT is a Cassegrain telescope with active control of the primary surface and focus, tilt, and lateral offset correction via a hexapod driven secondary. The Cassegrain beam is folded into the receiver cabin by a flat tertiary mirror. Utilising the full 50-m surface of the primary mirror gives an angular resolution for the LMT of $5.5\si{\arcsecond}$ at $\lambda = 1.1~\si{\milli\metre}$. After these three mirrors, which constitute the optical components of the LMT itself, MUSCAT uses an additional five mirrors to steer and focus the beam to the focal plane, as seen in Fig.~\ref{fig:Rays}.
\par
A planar fold mirror, M4, is switched to intercept the beam from M3 just ahead of the LMT's primary focus and selects MUSCAT as the active receiver, diverting the beam to MUSCAT's M5 mirror (an off-axis hyperboloid) and onto MUSCAT M6 (an off-axis paraboloid). M5 and M6 are arranged in a crossed-Dragone configuration, and form a demagnified image of the primary mirror vertically below M6. This pupil image is approximately $280~\si{\milli\metre}$ in diameter, and is used as a cold stop for the MUSCAT instrument. Within the MUSCAT instrument, the image of the telescope's prime focus is reformed by M7 (paraboloid) and M8 (hyperboloid) in another crossed-Dragone configuration. This produces an $f/3$ feed  across a $3.8\si{\arcminute}$ field of view; this is marginally reduced from the full $4\si{\arcminute}$ field of view of the LMT to reduce the possibility of spillover on the primary. Telecentricity and beam quality across the $138~\si{\milli\metre}$ diameter focal plane are excellent.
\par 
The MUSCAT operating spectral band (centred about $\lambda = 1.1~\si{\milli\metre}$) is defined through a filter stack within the MUSCAT instrument vessel. At the vacuum vessel, the optical port consists of an ultra-high-molecular-weight polyethylene (UHMWPE) window providing a hermetic seal with minimum distortion to the beam. A set of thermal-rejection filters mounted from the 50-kelvin stage remove the infrared load from the system and a set of low-pass edge metal-mesh filters at $13$, $12$, and $11~\si{\centi\metre^{-1}}$ define the upper frequency limit of the band (at the $3~\si{\decibel}$ point) as $306~\si{\giga\hertz}$. The lower frequency limit of the band will be defined by the waveguide section of the horn block (see Fig.~\ref{fig:pixelDesign}), initial simulations in ANSYS's HFSS software show the lower limit of the MUSCAT band is expected to be approximately $250\mbox{--}260~\si{\giga\hertz}$ however this will be confirmed during the final stages of detector development and testing.
\section{Cryogenic Design}
\begin{figure}[tb]
\centering
\includegraphics[height=0.4\textheight]{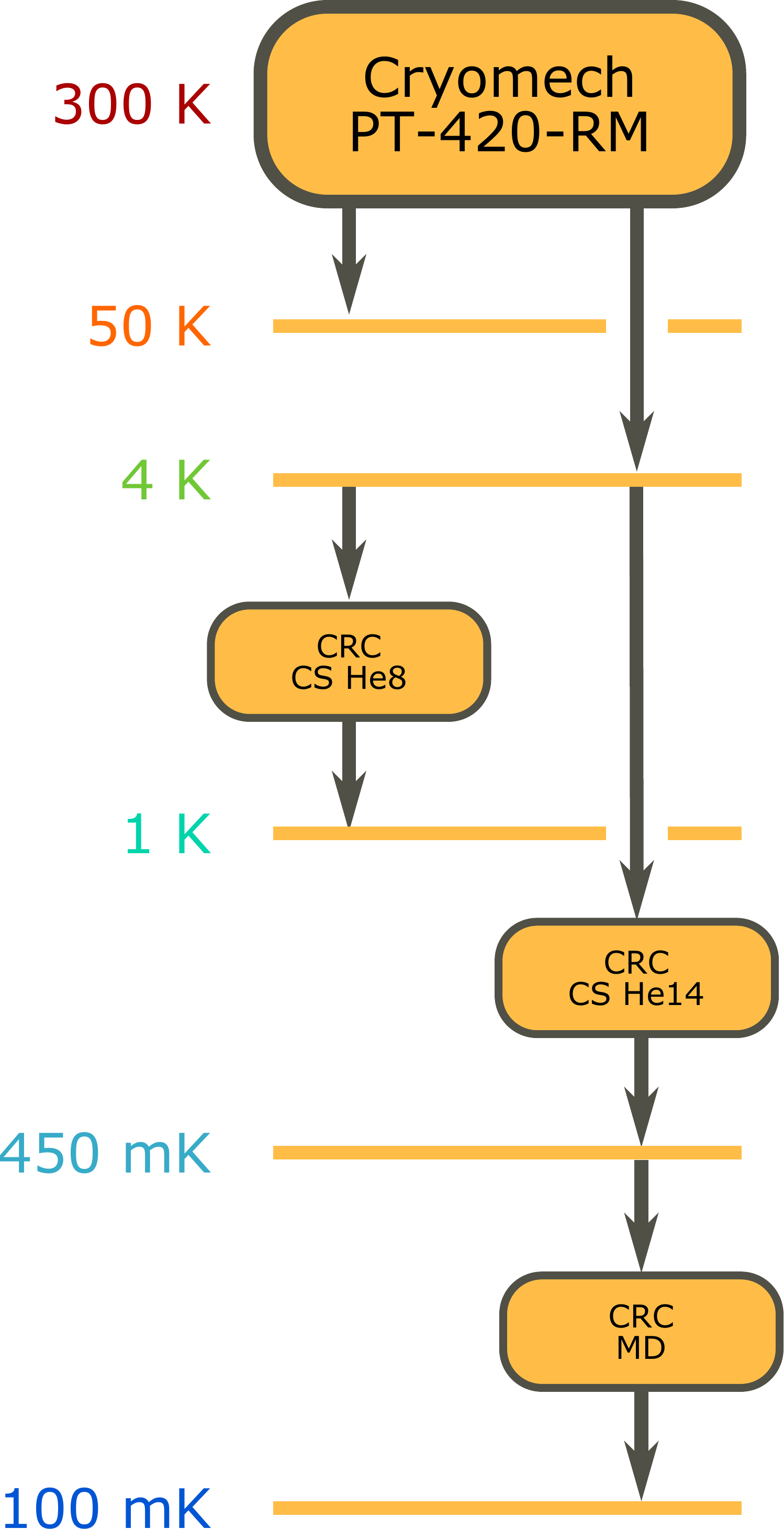}
\caption{The MUSCAT cooling system from room temperature ($300~\si{\kelvin}$) to the focal plane ($100~\si{\milli\kelvin}$).}\label{fig:coolingChain}
\end{figure}

MUSCAT will utilise a novel combination of cooling technologies to cool the detector array to of order $100~\si{\milli\kelvin}$ continuously. To make this possible, MUSCAT uses four separate refrigeration systems to cool the various stages of the instrument cryostat. These refrigerators are: a Cryomech\footnote{Cryomech Inc., 113 Falso Drive Syracuse, NY 13211, USA} PT-420-RM pulse tube cooler (PTC) used to cool the first two stages of MUSCAT to $50$ and $4~\si{\kelvin}$ respectively; two continuous sorption coolers (CS) developed by Chase Research Cryogenics\footnote{Chase Research Cryogenics, Neepsend Industrial Estate, 80 Parkwood Rd, Sheffield, S3 8AG, UK} (CRC) are then used to cool a heat sink stage to $\sim 1~\si{\kelvin}$ and a full stage and radiation shield to $450~\si{\milli\kelvin}$; finally, a miniature dilution refrigerator also developed by Chase Research Cryogenics cools the final stage---from which the focal plane is mounted---to $\sim 100~\si{\milli\kelvin}$. A simplified representation of the cooling chain used for MUSCAT is given in Fig.~\ref{fig:coolingChain}.
\par 
The two continuous sorption (CS) coolers used to cool the third and fourth stages of MUSCAT (see Fig.~\ref{fig:coolingChain}) are based upon the design described by Klemencic et al.\cite{Klemencic2016}. They each consists of a final head containing a small amount of \HeIII{} which acts as a ballast. This is cooled by a matched pair of precoolers which are cycled in anti-phase to each other. For the first stage cooled via a CS cooler, the precoolers simply consist of a single \HeIV{} sorption pump each, this is referred to as a He8 cooler (2 $\times$ \HeIV{}). The second stage cooled by a CS cooler uses a system where each precooler consists of a \HeIV{} pump and a \HeIII{} pump, referred to as a He14 cooler (2 $\times$ [\HeIV{}$+$\HeIII{}]). The operation of these units including how they are cycled in MUSCAT to accommodate large thermal loads, has been described in a previous paper.\cite{Brien2018}.
\par 
The final stage of MUSCAT, which contains the detector array, is cooled by a miniature dilution refrigerator. This system works on the same principle of forcing atoms of \HeIII{} across the boundary from a solution of pure \HeIII{} to dilute solution of \HeIV{} and \HeIII{}. A key feature of the design used here is that there is no active pumping of the gaseous \HeIII{} in the dilution circuit. This enables the entire dilution refrigerator to be contained within the cryostat and simplifies the operation. Under minimal loading these units have been shown to achieve a minimum temperature of approximately $50~\si{\milli\kelvin}$.\cite{Teleberg2008} In MUSCAT we anticipate achieving a temperature of order $100~\si{\milli\kelvin}$. This is limited by the load on our detector stage, which itself is dominated by the optical sky load which we expect to be $\sim 1.6~\si{\micro\watt}$.
\par 
Although four of MUSCAT's five refrigerators utilise \HeIII{}, which can be an expensive commodity costing upwards of 2,500 USD per S.T.P. litre of gas, the overall \HeIII{} requirement is only approximately $9~\si{\liter_{STP}}$. This enables higher-sensitivity detectors to be used with alower cost overhead from the cooling system and is made possible due to the relatively small volume of the miniature dilutor and the fact that the CS coolers do not require a long hold time to allow for regeneration of the \textit{standby} cooler. 
\section{Detector Design}
\begin{figure}[tb]
\centering
\subfloat[]{
\includegraphics[width=0.4\textwidth]{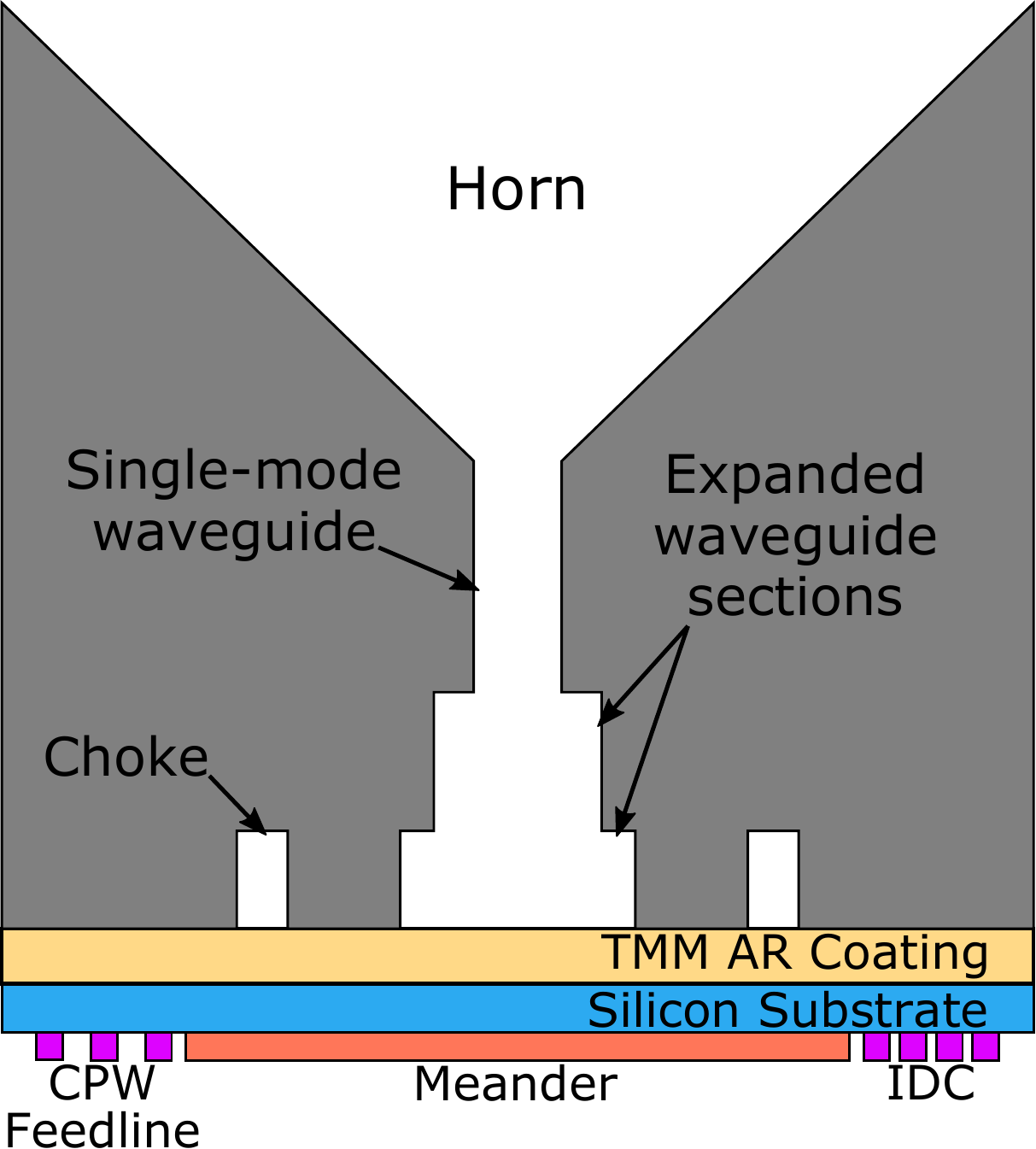}
\label{fig:pixelXsec}}
\hspace{0.1\textwidth}
\subfloat[]{
\includegraphics[width=0.4\textwidth]{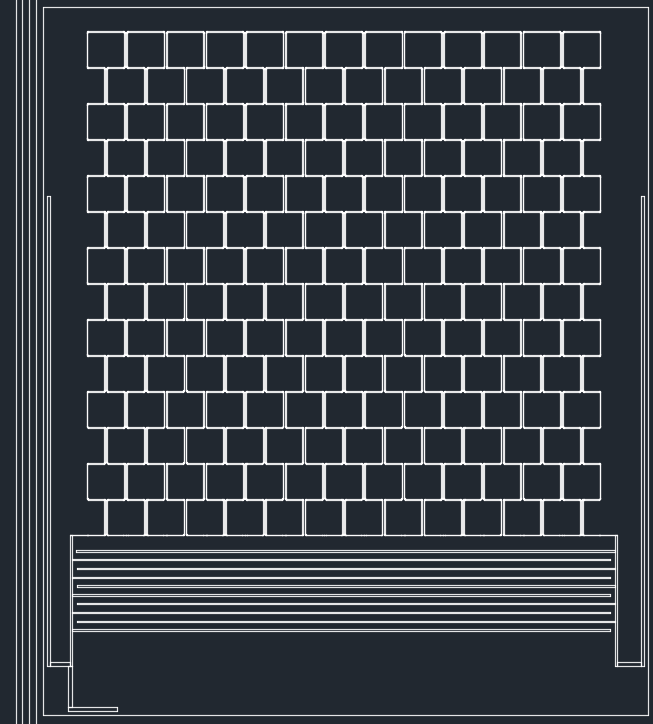}
\label{fig:pixel}}
\hspace{0.1\textwidth}
\caption{(a) Cross-sectional schematic of a single pixel on the MUSCAT focal plane. The absorbing meander of each LEKID is optically coupled via a single-moded horn and an interfacing layer of thermoset microwave material (TMM) acts as an anti-reflection coating. (b) Layout of a prototype pixel for MUSCAT showing the feed line to the left, the interdigitated capacitor (IDC) to the bottom and the meander structure at the top, for scale each finger of the IDC is $2.2~\si{\milli\metre}$ long.}\label{fig:pixelDesign}
\end{figure}
The MUSCAT focal plane will consist of approximately 1,500 horn-coupled LEKIDs. These will be placed across the focal plane using a hexagonal-packing arrangement with a spacing of $1F\lambda$. This layout, combined with horn coupling, provides good rejection to off-axis stray light and only illuminates the sensitive meander section removing the potential for light to scatter off non-sensitive elements such as the interdigital capacitor and feed-line and thus reduces the possibility of optical cross-talk in the array. A cross-sectional view of a single pixel is shown in Fig.~\ref{fig:pixelXsec}. Each pixel is rear-illuminated through a silicon substrate and to mitigate loss of response due to internal reflections, a layer of thermoset microwave material with refractive index $n=2.12$ (TMM4) is mounted between the horn block and the silicon wafer. The absorbing meander is fabricated from aluminium with a characteristic impedance of $1~\si{\ohm}/\square$. Expanded waveguide sections and RF chokes are used for impedance matching to the detector.
\par 
The LEKID architecture simplifies both design and fabrication by allowing optical power to be directly coupled to a highly responsive meander element without the need to use antennae or quasi-particle traps.\cite{Doyle2010} The first-generation focal plane on MUSCAT will not be polarisation sensitive, to realise this with a LEKID it is important to ensure that the inductive meander contains an equal length in each of the orthogonal directions. We have developed a meander geometry, shown in Fig.~\ref{fig:pixelDesign}, to meet this requirement. The total meander size is optimised for optical coupling and volume to provide high quality factor resonance features under the expected optical load. Maximising the resonator quality factor in this way minimises the potential for resonator-resonator clashes and improves the overall array yield. 
\par
\begin{figure}[tb]
\centering
\includegraphics[width=0.8\textwidth]{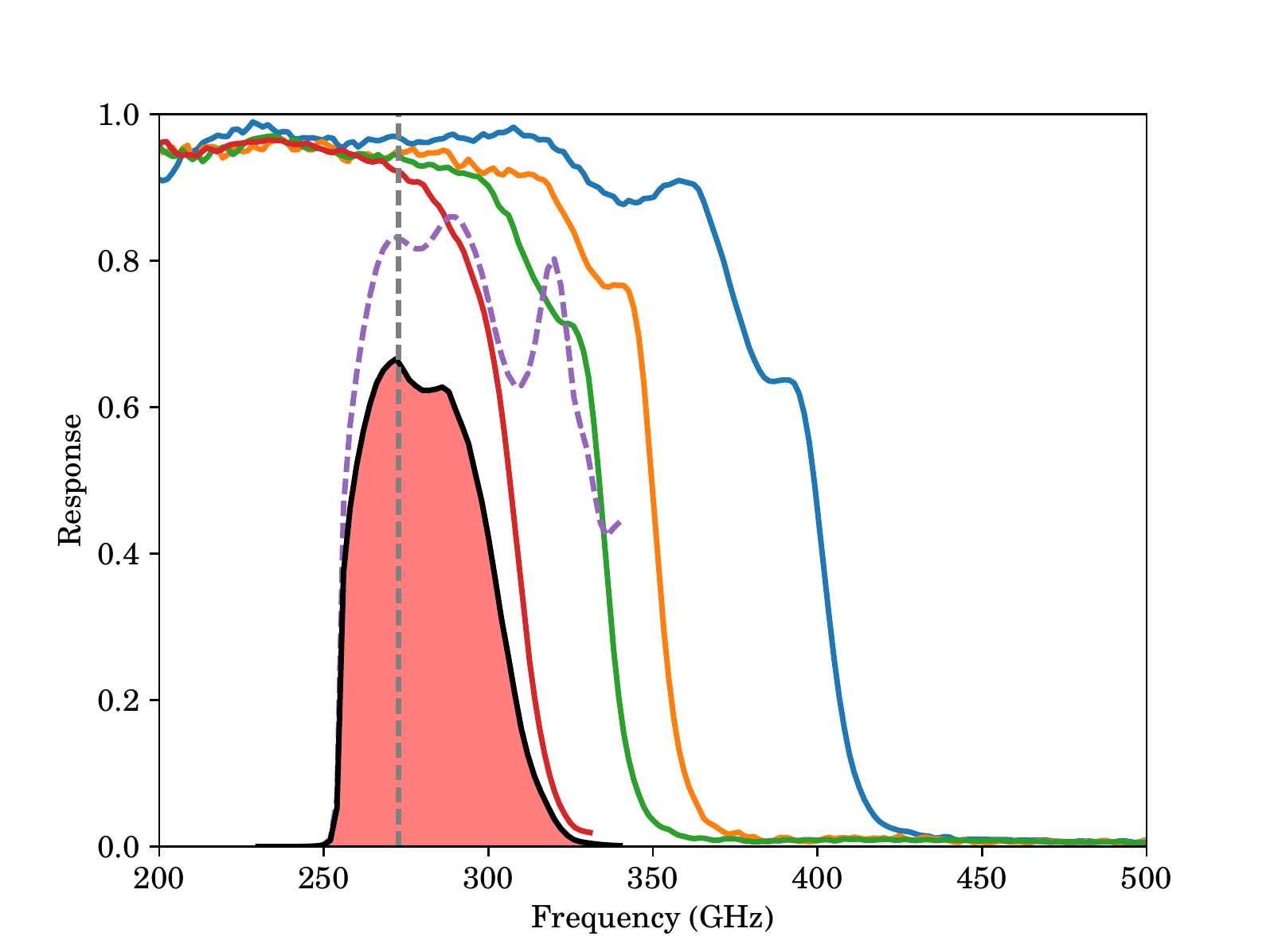}
\caption{Expected response of MUSCAT as defined by the metal-mesh filters (solid coloured curves) and HFSS modelling of the horn and waveguide (purple dashed line). These are combined to give the expected response band (black solid line and shaded area). The nominal $\lambda=1.1~\si{\milli\metre}$ ($\nu = 270~\si{\giga\hertz}$) band centre is show for illustration by the vertical grey dashed line.}\label{fig:MUSCATband}
\end{figure} 
To aid pixel design, detailed simulations of the expected absorption spectrum have been carried out using ANSYS's HFSS software. The results of this are given, for the described design with, in Fig.~\ref{fig:MUSCATband} which also shows the expected combined total response from both the horn and waveguide, and the metal-mesh filters. The detector shows a mostly broad-band response for frequencies between $260$ and $340~\si{\giga\hertz}$. The sharp cut on in response at $250~\si{\giga\hertz}$ is due to the horn's waveguide section (see Fig.~\ref{fig:pixelXsec}). 
\section{Readout}
The 1,500 pixels of the first-generation MUSCAT focal plane will be divided across six readout channels (MUX ratio of 250:1). Each cryogenic readout channel will consist of semi-rigid input and output coax cables. To balance minimising losses with thermal loading of the MUSCAT low-temperature stages, we have selected $2.15~\si{\milli\metre}$ outer diameter coax with a beryllium copper central conductor and a stainless steel outer. Thermal testing of these cables has been carried out and confirmed that they are compatible with the MUSCAT thermal budget however additional thermal protection will be implemented via the use of capacitive DC blocks with a high-pass frequency of $10~\si{\mega\hertz}$. The input into the detector blocks is attenuated at $4\mbox{-}\si{\kelvin}$ and $1\mbox{-}\si{\kelvin}$ stages to reduce the contribution of thermal noise in the probe signals. The output from each of the six sub arrays is passed through a cryogenic low-noise amplifier (LNA) developed by Arizona State University.\cite{Mani2014}. The LNAs have gain of approximately $30~\si{\decibel}$ and a noise temperature of $6~\si{\giga\hertz}$ throughout a $0.5\mbox{--}3.0~\si{\giga\hertz}$ readout band,
\par 
Probe tone generation and processing for each readout channel is handled by a ROACH-2 FPGA system\cite{Hickish2016} with MUSIC DAC/ADC boards. \cite{Duan2010} We will be using the firmware that has been developed for the BLAST-TNG instrument by Gordon et al.\cite{Gordon2016} The base band signals with $\pm 256~\si{\mega\hertz}$ bandwidth are mixed to and from the detector readout band ($\sim 0.6\mbox{--}1.1~\si{\giga\hertz}$) with a custom set of commercially available microwave electronic components.
\par
The demultiplexed detector data output is transferred by UDP stream to the Instrument Control System (ICS) computer at a rate of 488 samples per second per readout channel ($\sim 192~\si{\mebi\bit\per\second}$ total transmission bandwidth). The data packets are timestamped with values that are locked to one-pulse-per-second signal provided by the LMT for synchronisation with telescope other important data (such as telescope pointing and weather).
\par
Initial detector $IQ$ calibration will be performed by frequency sweeping of the local oscillator (LO) of each readout channel prior to beginning any observations. During observations, variations in detector response will be monitored by tracking the response to a modulated thermal source fed to the cold Lyot stop by a stainless steel light pipe from the outside of the instrument.
\section{Current Status}
At time of publication the mechanical design of MUSCAT has been finalised and the cryostat is in the final stages of manufacture with thermal characterisation partially complete to the 1-kelvin stage. The 450-mK stage---the final radiation shielding stage---is currently in fabrication and is expected to be integrated into instrument vessel in the coming months. 
\par 
A testing campaign exploring detector design and coupling optimisation has nearly been completed and we expect to finalise the detector design and move to array fabrication imminently.
\par 
The MUSCAT mirrors (M4--M8 in Fig.~\ref{fig:Rays}) are currently being manufactured by companies in Mexico. The final two mirrors are mounted inside the instrument vessel and will be shipped to Cardiff, UK for system integration and testing. The remaining mirrors will be shipped directly to the LMT for MUSCAT's commissioning.
\section{Conclusion}
MUSCAT is currently nearing the end of its construction and should be shipping to the LMT in late 2018. For its first on-sky run, MUSCAT will utilise 1,500 LEKIDs read out using six readout channels with continuous cooling of the focal plane to approximately $100~\si{\milli\kelvin}$. During this first observing semester MUSCAT will help to allocate optical counterparts to sources in the \textit{Herschel}-ATLAS catalogue and map star forming regions beyond $400~\si{\parsec}$. MUSCAT has been design to be a versatile instrument with the capability of being upgraded with relative ease to act as a testbed for new technologies.

\section*{Acknowledgements}
This work has been supported by Research Councils UK (RCUK) and Consejo Nacional de Ciencia y Tecnolog\'{i}a (CONACYT) under the Newton Fund, project ST/P002803/1. MUSCAT has also been supported by Chase Research Cryogenics Ltd and Xilinx.
\bibliography{bib}
\bibliographystyle{spiebib}
\end{document}